\documentclass[12pt]{iopart}

\usepackage{iopams}
\usepackage{cite}
\usepackage{graphicx}
\usepackage{enumerate}
\usepackage{color}
\usepackage{url}
\usepackage[usenames,dvipsnames]{xcolor}
\usepackage{braket}
\usepackage{siunitx}

\usepackage{multirow}
\begin{document}

\title[]{VUV diagnostic of electron impact processes in low temperature molecular hydrogen plasma}

\author{J. Komppula$^1$ and O. Tarvainen$^1$}

\address{$^1$ University of Jyvaskyla, Department of Physics, P.O. Box 35, FI-40014 University of Jyvaskyla, Finland}
\ead{jani.komppula@jyu.fi}

\begin{abstract}
Novel methods for diagnostics of molecular hydrogen plasma processes, such as ionization, production of high vibrational levels, dissociation of molecules via excitation to singlet and triplet states and production of metastable states, are presented for molecular hydrogen plasmas in corona equilibrium. The methods are based on comparison of rate coefficients of plasma processes and optical emission spectroscopy of lowest singlet and triplet transitions, i.e. Lyman-band ($B^1\Sigma^+_u \rightarrow X^1\Sigma^+_g$) and molecular continuum ($a^3\Sigma^+_g \rightarrow b^3\Sigma^+_u$), of the hydrogen molecule in VUV wavelength range. Comparison of rate coefficients of spin-allowed and/or spin-forbidden excitations reduces the uncertainty caused by the non-equilibrium distributions of electron energy and molecular vibrational level, which are typically known poorly in plasma sources. The described methods are applied to estimate the rates of various plasma processes in a filament arc discharge.
\end{abstract}

\submitto{\PSST}
\maketitle
\section{Introduction}

Molecular hydrogen plasmas play an important role in negative (H$^-$) \cite{Bacal_volume_production} and molecular (H$^{+}_{2}$) hydrogen \cite{H_2_plus_source_1,H_2_plus_source_2,H_2_plus_source_3} ion sources. Furthermore, hydrogen molecules are an important species in magnetically confined plasmas of fusion devices \cite{H_2_fusion} as well as in astrophysics \cite{H_2_astrophysics}.

Optical emission spectroscopy (OES) is a powerful and widely used plasma diagnostic method. Wavelength range from near ultraviolet to near infrared is often used for the purpose (see e.g. \cite{allHydrogenEmissionSpectroscopy} and references therein). Most of the diagnostic methods based on OES are developed to obtain information about plasma parameters, such as the plasma density and temperature, which requires employing sophisticated collisional radiative models. Very few methods based on molecular emission have been developed for diagnostics of hydrogen plasma processes. For example, Sawada and Fujimoto \cite{molecular_visible_diagnostics1} have developed methods for determining ionization and dissociation rate coefficients of hydrogen molecules based on optical emission spectroscopy of hydrogen atoms in visible light range. Graham has estimated the production rate of ground state hydrogen molecules at high vibrational levels based on (absolute) Lyman-alpha emission and comparison of the rate coefficients of molecular (singlet) and atomic excitations \cite{Graham_1984_VUV_measurement}. Lavrov et al. have demonstrated that molecular continuum emission could be used as a probe for dissociation rate in molecular hydrogen plasma \cite{Lavrov_1999_continuum}. However, only excitations to $a^3\Sigma^+_g$ and $b^3\Sigma^+_u$ states were taken into account and the total calculated emission based on an extrapolation of spectral measurement data in the range of 225-450~nm.

Measurement of the light emission in the VUV wavelength range (100--250 nm) is a relevant optical diagnostic of certain plasma processes in molecular hydrogen plasmas.  This is because the transitions from the lowest excited electronic states to the ground state emit photons in this range, radiative lifetimes of the lowest excited states are short and electron impact excitation cross sections from the ground state to those states are approximately an order of magnitude greater than excitation cross sections to upper electronic states. Therefore, the physical origin of the population density of the lowest excited states can be understood straightforwardly and photon emission rate is linearly proportional to the electron impact excitation rate from the ground state.

Application of VUV-emission for diagnostics of molecular hydrogen plasma processes such as ionization, dissociation and excitation, is demonstrated in this paper. Instead of using information about the plasma parameters i.e. density and temperature, the applied method is based on direct measurement of the VUV-emission rate, which is proportional to the electron impact excitation rate. Comparison of the rate coefficients results to a robust plasma diagnostic method requiring minimum knowledge on the plasma parameters. The method can be applied to obtain absolute numbers or relative changes of volumetric production rates through absolute or relative measurements of the VUV-irradiance. The error caused by the electron energy distribution (EED) and molecular vibrational temperature is discussed.

The presented methods are applied for a set of previously presented VUV-irradiance measurements \cite{komppula_NIBS_2012} of a filament driven negative hydrogen ion source, LIISA, at the JYFL accelerator laboratory. Estimated excitation and ionization rates based on saturation value (at \textgreater 100 eV electron energy) of rate coefficients of vibrationally cold plasma, have been presented and discussed briefly in the earlier study. In this study the analysis is carefully extended to account for different EEDF's and vibrational temperatures. Furthermore, the dissociation rate of hydrogen molecules and the production rate of metastable hydrogen molecules ($c^3\Pi_u$ state) are determined for the first time.

\section{Processes affecting the population densities of excited electronic states of hydrogen molecules}
\label{sec:population_density}

Hydrogen molecule has two multiplet systems of electronic states, singlet and triplet, which differ by the orientation of the electron spins ($S$=0 for singlet and $S$=1 for triplet) governed by the Pauli exclusion principle. Radiative transitions between the electronic states are classified as optically allowed and optically forbidden. Optically allowed electron transitions are subject to the following selection rules: (a) change of the parity of the total wave function ($g\leftrightarrow u$), (b) conservation of the total spin quantum number i.e. $\Delta S$=0 and (c) possible change of the total angular momentum quantum number $\Delta\Lambda$=0$,\pm 1$. 

\begin{figure}
 \begin{center}
 \includegraphics[width=0.45\textwidth]{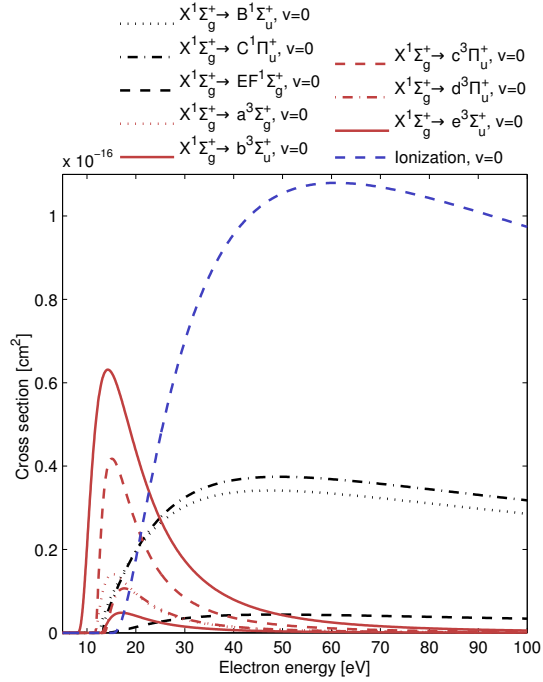}% % Important NOTE: Please make certain your figures do not %include local directory paths. ex. "c:\file\sub\fig1.eps"
 \caption{\label{fig:cross_sections} The cross sections of the most significant electron impact processes of hydrogen molecules  ($\nu=0$) with hot electrons ($E_e>$ 9 eV). Excitations from the ground state to singlet states are marked with black color, to triplet states with red color and ionization with blue color. The cross sections are from Ref. \cite{Janev2003}.}%
 \end{center}
 \end{figure}

The selection rules affect the functional shapes of electron impact excitation cross sections. However, due to electron exchange effects in electron-impact excitation collisions at low energies, these selection rules are not strictly preserved\cite{Janev2003}. The most significant differences of electron impact cross sections are found between singlet-triplet ($\Delta S$=1) spin-forbidden excitations and singlet-singlet ($\Delta S$=0) spin-allowed excitations. The excitation cross sections from the ground state ($X^1\Sigma^+_g$) to other singlet states ($B^1\Sigma^+_u$, $C^1\Pi_u$, $B'^1\Sigma_u$, $D^1\Pi_u$, etc.) are rather insensitive to the impacting electron energy in the range \textgreater30~eV (Fig. \ref{fig:cross_sections}). In the case of spin-forbidden transitions ($\Delta S$=1) the electron impact excitation occurs via resonance processes. Therefore, the electron impact excitation cross sections from the ground state ($X^1\Sigma^+_g$) to  the triplet states ($b^3\Sigma^+_u$, $c^3\Pi_u$, $a^3\Sigma^+_g$, etc.) are peaked at \textless20~eV (Fig. \ref{fig:cross_sections}). 

Altogether the hydrogen molecule has several tens of electronic states \cite{H_Franck-Condon_factors}. In this study singlet states $B^1\Sigma^+_u$, $C^1\Pi_u$, $EF^1\Sigma_g$ are taken into account. This is because the electron impact excitation cross sections of  $B^1\Sigma^+_u$ and $C^1\Pi_u$ states correspond to over 75\% of the total electron impact excitation cross section to singlet states and cascading from the $EF^1\Sigma_g$ state is the only significant effect involving upper states. Other singlet states either emit VUV-light in different wavelength range (e.g. $D^1\Pi_u$ and $B'^1\Sigma_u$)  or the corresponding electron impact excitation cross sections are orders of magnitude lower (e.g. $B''^1\Sigma_u$, $D'^1\Pi_u$, $GK^1\Sigma_g$) \cite{Janev2003}. 

On the other hand, triplet states $a^3\Sigma^+_g$, $b^3\Sigma^+_u$, $c^3\Pi_u$, $d^3\Pi_u$ and $e^3\Sigma^+_u$ are taken into account in this study. This is due to lack of available cross sections for electron impact excitations to other triplet states. The lowest triplet state ($b^{3}\Sigma_{u}^{+}$) to which the radiative decay chains of $a^3\Sigma^+_g$, $e^3\Sigma^+_u$ and $d^3\Sigma^+_g$ states end, is repulsive. The $c^3\Pi_u$ state is partially metastable depending on the symmetry ($c^3\Pi_u^+$ or $c^3\Pi_u^-$) \cite{H2_c_lifetime_calculation}. The radiative lifetime of the metastable $c^3\Pi_u^-$ state depends on the vibrational level and ranges from \SI{1}{\milli\second} ($\nu$=0) to \SI{10}{\micro\second} ($\nu$=3) \cite{metastable_calculation_Freis,statelifetimes}. The $c^3\Pi_u$ state is sensitive to electron impact de-excitation/ionization \cite{transitions_triplet_1,transitions_triplet_2} and collisional quenching \cite{H2_c_quenching}. 

Plasma emission spectroscopy yields information about population densities of molecules on excited states, which are affected by several processes. The diagnostics presented in this study are based on the premise, that the VUV-light emission rate is equal to the electron impact excitation rate from the ground state to the corresponding electronic states (corona model). This sets limitations on the temporal resolution of the diagnostic as well as plasma parameters, which are fullfilled by most low temperature laboratory plasmas including hydrogen ion sources (see e.g. Refs. \cite{plasma_sources,Bacal_volume_production} and references therein):
\begin{enumerate}[i)]
\item\label{item:time_scale}The shortest temporal scale that can be studied is on the order of 1 ns for singlet transitions and 100~ns for triplet transitions
\item\label{item:electron_density} The plasma density is less than $10^{14}$~cm$^{-3}$
\item\label{item:neutral_pressure} The neutral gas pressure is under 500~Pa
\item\label{item:ion_temperature} The ion temperature $T_i$ is lower than the electron temperature $T_e$
\end{enumerate} 
The listed limitations do not concern the metastable $c^3\Pi_u$ state, which is discussed thoroughly in Sections \ref{cap:metastable} and \ref{cap:dissociation}

The best possible temporal resolution of the diagnostic method is limited by the radiative lifetime of the excited states (delay in spontaneous emission). The radiative lifetimes of the lowest singlet states ($B^1\Sigma^+_u$ and $C^1\Pi_u$) and lowest triplet states are under 1~ns and 10--40~ns respectively \cite{statelifetimes}.

The plasma density is limited by collisional de-excitation and ionization rate from the excited states. Significant collisional de-excitation of $B^1\Sigma^+_u$ state requires the electron density to be $10^{14}$--$10^{16}$~cm$^{-3}$ \cite{transitions_singlet}. For the $a^3\Sigma^+_g$ state the corresponding number is $10^{15}$~cm$^{-3}$ \cite{transitions_triplet_1}. Since the cross sections of electron-impact ionization from the excited states ($B^1\Sigma^+_u$ and $a^3\Sigma^+_g$) are on the same order of magnitude (as a maximum) with the electron impact de-excitation cross sections \cite{Janev2003}, the given maximum plasma densities are valid also for ionization. 

Electronic states can also be de-excited e.g. via excitation transfer with neutral particles, charge exchange with ions and penning ionization. The cross sections of these processes are, however, known poorly \cite{Janev2003}. Most of these de-excitation processes are resonant by nature and hence their cross sections could be large, but the required collision energies are on the order of a few eV. Thus, they are negligible in low temperature plasmas, where the the average energies of neutral particles and ions are low. The excitation transfer with neutral particles could limit the maximum neutral gas pressure. However, there are no cross section data available for excitation transfer of hydrogen molecule. The lower limit for the maximum neutral gas pressure ($T$=300~K) can be estimated by using the excitation transfer cross section of hydrogen atom (approximately $10^{-13}$~cm$^2$), which yields a pressure of 5~mbar assuming 10~ns lifetime for the excited state. The given cross section is probably several orders of magnitude too large (and calculated upper limit of the pressure too low), because of the vibrational distribution of the molecules and the Franck-Condon principle decreasing the excitation transfer probability of the molecule in comparison to the atom.

The applied model requires that the electron impact excitation is the dominant excitation process. This in turn requires that the ion temperature is lower than the electron temperature. The effects of photonic excitation and recombination on the population densities of excited molecules are negligible in comparison to corresponding effects in the case of atomic emission. Molecular hydrogen (laboratory) plasmas are optically thin for molecular emission, because the vibrational distribution of molecules and Franck-Condon principle result to low probability of photon excitation. On the other hand, recombination of the hydrogen molecule leads mainly to dissociation of the molecule \cite{Janev2003}, i.e.
\begin{equation}
e+H_2^+ \rightarrow \{H_2^{**};H_2^{*Ryd}\}\rightarrow H(1s)+H(n\geq2),
\end{equation}
which does not contribute to the population densities of excited molecules.

\section{Estimating rate coefficients from VUV-emission}
 \label{sec:diagnostics}
\begin{figure}
 \begin{center}
 \includegraphics[width=0.45\textwidth]{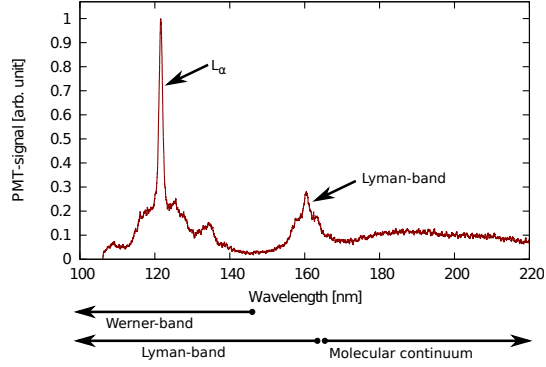}% % Important NOTE: Please make certain your figures do not %include local directory paths. ex. "c:\file\sub\fig1.eps"
 \caption{\label{fig_spectra} VUV emission spectrum of a hydrogen plasma of an arc discharge with 840 W discharge power and 0.5 Pa pressure. The spectrum is not correlated for spectral transmittance.}%
 \end{center}
 \end{figure}
 
A typical VUV spectrum of a hydrogen plasma is presented in Fig. \ref{fig_spectra}. The spectrum can be divided into four different regions based on the origin of the radiation. Lyman-series (mainly Lyman-alpha) radiation is emitted by hydrogen atom while three dominant molecular transitions emit VUV-light. Those are Lyman- and Werner-bands, corresponding to transitions from the lowest excited singlet states ($B^1\Sigma^+_u$ and $C^1\Pi_u$) to the ground state, and molecular continuum corresponding to the lowest transition between triplet states (a$^3\Sigma^+_g \rightarrow $b$^3\Sigma^+_u$). Because the lowest triplet state is repulsive, a continuum is observed instead of a band structure.

The Lyman-band and molecular continuum emissions are the most suitable transitions for the purpose of plasma diagnostics. This is because significant parts of Lyman-band and molecular continuum radiation are emitted in ranges of 145--165~nm and 170--240~nm including negligible contribution from other transitions. Furthermore, structure of Lyman-band and molecular continuum emissions do not depend (significantly) on the plasma parameters and cascade effects from the upper states to $B^1\Sigma^+_u$ state are well known, which allows straightforward estimation of total emission from the measured signals in the specific ranges.

VUV-emission can be used as a probe for the most important plasma processes, such as ionization rate, vibrational excitation rate to high vibrational levels, production rate of metastable $c^3\Pi_u$ states, molecular dissociation rate and hot electron density. These rates are sensitive to distributions of electron energy and molecular vibrational level.

The vibrational distribution of neutral hydrogen molecules is not in the thermal equilibrium, especially at high vibrational levels. However, a majority (\textgreater90\%) of neutral molecules in typical laboratory plasmas are on the lowest vibrational levels ($\nu\leq4$) (e.g. Refs. \cite{experimental_vibrational_temperature, the_simulation_paper} and references therein). This part of the distribution can be described reasonably well with the Boltzmann distribution, i.e. using a vibrational temperature $T_{vib}$. The vibrational temperature of the lowest vibrational levels has been found to be between 100~K and 10000~K (e.g. Refs. \cite{the_simulation_paper,FantzMolC} and references therein). These numbers are used as extremes when estimating the uncertainty of the presented diagnostic caused by the vibrational distribution. This is because the cross sections of electron impact ionization and excitation to electronic states are less than linearly proportional to the vibrational quantum number. Molecules have also rotational levels, which have a negligible effect on the discussed electron impact processes due to minimal energy exchange ($\approx$0.01~eV) in transitions and long collision times (“frozen rotation”) \cite{Janev2003}.

The electron energy distribution functions (EEDF) of low temperature plasmas are typically non-Maxwellian. The distributions are often bi-Maxwellian at low pressures and Druyvesteyn-like at high pressures \cite{the_book}. At high plasma density (and ionization degree) electron-electron collisions tend to drive the EEDF towards a Maxwellian \cite{the_book}. Furthermore, superpositions of Maxwellian and flat (see definition later) distributions have been commonly found in filament arc discharges (e.g. Refs. \cite{flat_EEDF_1,flat_EEDF_3,flat_EEDF_2}). When electron-electron collisions dominate in the entire energy range, the resulting EEDF is Maxwellian by definition. In hydrogen plasmas this is the situation only at low electron energies ($E_{e}$\textless9~eV). The EEDF at high electron energies is determined by elastic and inelastic collisions, plasma heating method and electron confinement.

Typically only high energy electrons ($E_e$\textgreater20~eV) contribute to ionization and excitation to singlet states. For example, in the case of the EEDFs reported in Refs. \cite{Graham_1995_kinetics_of_negative_hydrogen_ions,Sun_2010_RF_H2_PIC_simulation,McNeely_2008_Langmuir_probe_from_batman}, \textgreater 20~eV electrons contribute more than 99\% to the ionization and 76--91\% to the excitation to $B^1\Sigma^+_u$ state although their density is only 1--3\%  of the total electron density in Ref. \cite{Graham_1995_kinetics_of_negative_hydrogen_ions,Sun_2010_RF_H2_PIC_simulation} or 30\% in Ref. \cite{McNeely_2008_Langmuir_probe_from_batman}. Therefore, ionization and singlet state excitation rates can be described reasonably accurately by focusing only to the high energy part of the EEDF. Excitations to triplet states are most sensitive to the electron energy in the range of 10--20~eV.

\begin{figure}
 \begin{center}
 \includegraphics[width=0.45\textwidth]{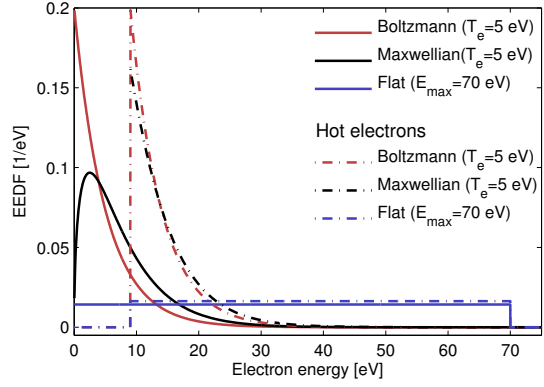}% % Important NOTE: Please make certain your figures do not %include local directory paths. ex. "c:\file\sub\fig1.eps"
 \caption{\label{fig:EEDF} Examples of electron energy density functions used in this study. Hot electron versions of distributions are used for calculating the rate coefficients presented in Section \ref{cap:hot electron density} by renormalizing the $E_{e}>$  9 eV part. The choice of 9 eV corresponds to the threshold energy of $X^1\Sigma^+_g \rightarrow B^1\Sigma^+_u$ electron impact excitation for vibrational levels $\nu\leq4$.}%
 \end{center}
 \end{figure}

In this paper two EEDFs have been chosen for a closer study from a mathematical point of view to describe the distribution of electrons at energies exceeding the excitation and ionization thresholds (Fig. \ref{fig:EEDF}). At high energies the Maxwellian distribution and the tail of the bi-Maxwellian distribution can be described by the Boltzmann probability function 
\begin{equation}
f(E_e)=Ae^{-E_e/k_bT_e},
\end{equation}
where $A$ is a normalization factor, $E_e$ is electron energy, $k_b$ is the Boltzmann constant and $T_e$ is electron temperature. The high-energy tail of the EEDF in filament arc discharges can be described by a flat distribution of energies, 
\begin{equation}
f(E_e)=\left\{
  \begin{array}{lr}
    A & : E_e \leq E_{max}\\
    0 & : E_e > E_{max},
  \end{array}
\right.
\end{equation}
where $A$ is a normalization factor and $E_{max}$ is the end point (or maximum) energy of the plasma electrons corresponding to the potential difference between the cathode and the plasma. The rate coefficients of these two distributions differ significantly. If the average energy of the distribution ($\frac{3}{2}k_bT_e$ for Maxwellian distribution and $E_{max}$/2 for the flat distribution) changes, the electron density at energies corresponding to the most sensitive range of the rate coefficients (6~eV\textless$E_e$\textless40~eV, see discussion in Sections \ref{cap:ionization}--\ref{cap:dissociation}) changes exponentially in the case of Maxwellian distribution and lineraly in the case of the flat distribution. 

The VUV-emission rate corresponds to the excitation rate of neutral molecules, if the plasma parameters are within the limits specified in section \ref{sec:population_density}. Comparison of the rate coefficients of different processes allows estimating their volumetric reaction rates from the measured VUV-emission rates of specific emission bands. The volumetric rate $R$ of an electron impact process can be described as
\begin{equation}
\label{eq:reaction_rate}
R= n_e n_n \int f(v) v \sigma(v) dv=n_nn_e\braket{v\sigma},
\end{equation}
where $n_e$ is the electron density, $n_n$ the neutral molecule density, $f$ normalized electron velocity distribution function (EVDF), $v$ the electron velocity and $\sigma$ the cross section of the process. The term $\braket{v\sigma}$ is the rate coefficient which depends on the functional shapes of the EVDF and the cross section.

In molecular plasmas the volumetric rate of any given process depends on the vibrational distribution. The total volumetric rate can be calculated as a weighted average of the rate coefficients corresponding to individual vibrational levels. For Boltzmann distribution the result can be written as
\begin{equation}
\label{eq:reaction_rate_vibration}
R(T_{vib})=n_nn_e\frac{\sum_i \braket{v\sigma}_i \exp(-\frac{E_i}{kT_{vib}})}{\sum_i \exp(-\frac{E_i}{kT_vib})}=n_nn_e \alpha(T_{vib}),
\end{equation}
where $T_{vib}$ is the vibrational temperature, $E_i$ is the energy of the vibrational level $i$ on the ground state and $\braket{v\sigma}_i$ is the total excitation rate coefficient from the ground state vibrational level $i$ to all possible vibrational levels of the upper  state.

If the volumetric rate of a specific process ($R_1$) can be determined e.g. by measuring the volumetric emission rate of a specific band of VUV-light, the volumetric rate of another process ($R_2$) can be estimated from
\begin{equation}
\label{eq:ratio_of_rate_coefficients}
R_2=\frac{n_{n}n_e\alpha_2(T_{vib})}{n_{n}n_e\alpha_1(T_{vib})} R_1.
\end{equation}
If the same neutral species ($n_{n}$) is involved in both processes, the electron and neutral densities cancel out. Thus, the volumetric rate $R_2$ can be obtained by multiplying the measured volumetric rate $R_1$ with the corresponding rate coefficients, i.e.
\begin{equation}
\label{eq:ratio_of_rate_coefficients_simpler}
R_2=\frac{\alpha_2(T_{vib})}{\alpha_1(T_{vib})} R_1=K_{2,1}(T_{vib}) R_1,
\end{equation}
where $K_{2,1}$ is the ratio of the rate coefficients. 

It is of note that the ratio of the rate coefficients, $K_{2,1}$, is significantly less sensitive to changes of the EEDF (e.g. in terms of average electron energy or functional shape), and often to the vibrational temperature, than individual rate coefficients.  The sensitivity of the $K_{2,1}$ to the EEDF can be understood by studying the cross sections of the corresponding processes (Fig. 1) and their ratios at different electron energies. The rate coefficient ratio is sensitive to changes of the EEDF in the energy range where both, the ratio of the cross sections changes significantly, and the absolute values of the cross sections are large. Typically the sensitive energy range is a narrow band close to the threshold energies. For example, when the excitation rate coefficients of similar types of transitions (in terms of the selection rules) are compared the sensitive energy range is 10--20~eV. In such cases (similar types of transitions) the vibrational temperature dependence of $K_{2,1}$ is reduced due to similar dependence of the individual cross sections on it. Therefore, in some cases $K_{2,1}$ is insensitive to the average electron energy, shape of EEDF and vibrational temperature although individual rate coefficients can vary by orders of magnitude within the same ranges of EEDF and vibrational temperature variations. Such situations are presented in Sections 3.2--3.6, where the extreme values of $K_{2,1}$ vary only by \textless30\% in comparison to their average values in the entire energy range of (both) EEDFs and vibrational temperatures.

There are several reviews about electron impact cross sections in low temperature hydrogen plasmas (e.g. Refs. \cite{Janev2003,crosssections1,crosssections2,Celiberto_2001_cross_section_functions,Shakhatov_2009_Cross_section_database}). Data from Ref. \cite{Janev2003} has been used in this study because it includes the most complete critically assessed set of cross sections including fitting functions in analytic form.
 
\subsection{Molecular ionization rate}
\label{cap:ionization}

\begin{figure}
 \begin{center}
 \includegraphics[width=0.45\textwidth]{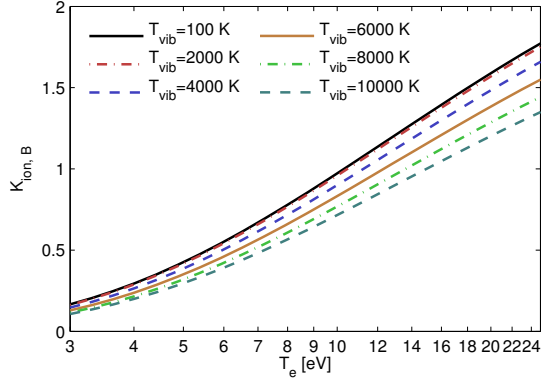}% % Important NOTE: Please make certain your figures do not %include local directory paths. ex. "c:\file\sub\fig1.eps"
 \caption{\label{fig:ionization_MBD} The ratio $K_{ion,B}$ of ionization and $X^1\Sigma^+_g \rightarrow B^1\Sigma^+_u$ excitation rate coefficients as a function of the temperature ($T_{e}$) of a Maxwellian EEDF and vibrational temperature of the hydrogen molecules. Calculated using cross sections from Ref. \cite{Janev2003}.}%
 \end{center}
 \end{figure}

\begin{figure}
 \begin{center}
 \includegraphics[width=0.45\textwidth]{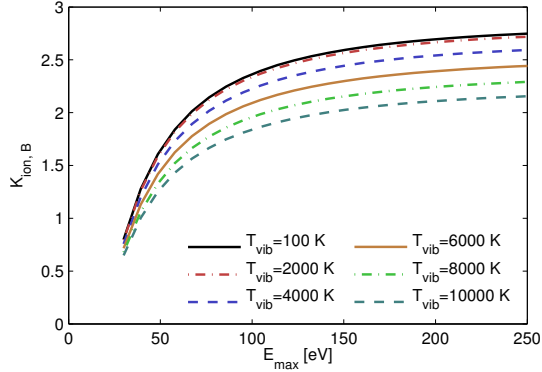}% % Important NOTE: Please make certain your figures do not %include local directory paths. ex. "c:\file\sub\fig1.eps"
 \caption{\label{fig:ionization_FD} The ratio $K_{ion,B}$ of ionization and $X^1\Sigma^+_g \rightarrow B^1\Sigma^+_u$ excitation rate coefficients as a function of the maximum energy ($E_{max}$) of a flat EEDF and vibrational temperature of the hydrogen molecules. Calculated using cross sections from Ref. \cite{Janev2003}.}%
 \end{center}
 \end{figure}

Hydrogen molecules can be ionized by electron impact non-dissociatively via $H_2^+(X^2\Sigma_g^+;\nu')$ state and dissociatively via $H_2^+(X^2\Sigma_g^+;\epsilon')$ and $H_2^+(B^2\Sigma_u^+;\epsilon)$ states \cite{Janev2003}. The cross-section of non-dissociative ionization is more than an order of magnitude larger ($\nu=0$) than the total cross section of the dissociative ionization \cite{Janev2003}. Hence, only non-dissociative ionization is taken into account in this study. The functional shapes of electron impact cross sections of ionization and spin-allowed excitations to the electronic states are similar. The molecular ionization rate can be estimated from the measured Lyman band emission rate with Eq. \ref{eq:ratio_of_rate_coefficients_simpler} by estimating the ratio of the ionization and $X^{1}\Sigma_{g}^{+} \rightarrow B^{1}\Sigma_{u}^{+}$ excitation rate coefficients, $K_{ion,B}$, plotted for different EEDFs and vibrational temperatures in Figs.  \ref{fig:ionization_MBD} and \ref{fig:ionization_FD}. 

In the case of $K_{ion,B}$, the most sensitive range is 10--40~eV. The sensitivity in this range is mostly due to different threshold energies of the processes; 11.6~eV for $B^{1}\Sigma_{u}^{+}$-excitation and 15.4~eV for ionization. In the case of Maxwellian EEDF, there is a significant plasma temperature dependence of the rate coefficient ratio due to exponential nature of the EEDF in the range of 10--40~eV. In the case of the flat EEDF, the ratio of the rate coefficients does not depend strongly on the maximum energy of the distribution if $E_{max}$\textgreater50~eV, which is typical in filament driven arc discharges.

It can be concluded from Figs.  \ref{fig:ionization_MBD} and \ref{fig:ionization_FD}, that the uncertainty of $K_{ion,B}$ caused by the vibrational temperature is \textless$\pm15$\%, if $T_{vib}$=6000~K is used and the real vibrational temperature is in the range of 100--10000~K. If the electron temperature can be estimated (or measured) with an accuracy of $\pm20$\% which can be considered as a typical number \cite{McNeely_2008_Langmuir_probe_from_batman}, it causes an uncertainty of \textless$\pm20$\% to $K_{ion,B}$. Hence, the total uncertainty of $K_{ion, B}$ is \textless$\pm25$\% in the case of Maxwellian EEDF. It must be emphasized that the flat distribution does not accurately describe the real EEDF. However, the uncertainty of the diagnostics results caused by such deviation can be described as a variation of (the effective) $E_{max}$ of an ideal flat EEDF. This can be demonstrated as follows: (1) choosing $K_{ion,B}=2$ corresponds to an uncertainty of \textless$\pm$25\% for 60~eV\textless$E_{max}$\textless250~eV or (2) choosing $K_{ion,B}=1.5$  corresponds to an uncertainty of \textless$\pm$35\% for 35~eV\textless$E_{max}$\textless85~eV.
 
\subsection{Excitation rate to $B^1\Sigma^+_u$ and $C^1\Pi_u$ states}
\label{cap:vibrational_excitation}

\begin{figure}
 \begin{center}
 \includegraphics[width=0.45\textwidth]{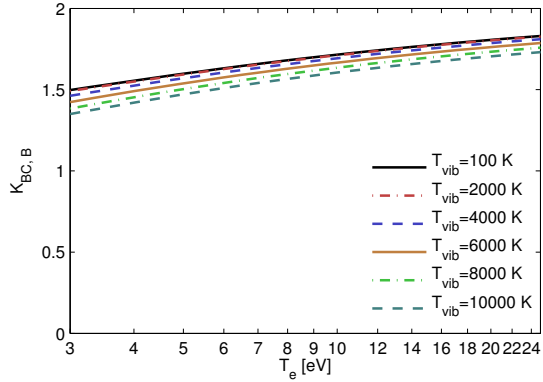}% % Important NOTE: Please make certain your figures do not %include local directory paths. ex. "c:\file\sub\fig1.eps"
 \caption{\label{fig:BC_MBD} The ratio $K_{BC,B}$ of $X^1\Sigma^+_g \rightarrow B^1\Sigma^+_u, C^1\Pi_u$ and $X^1\Sigma^+_g \rightarrow B^1\Sigma^+_u$ excitation rate coefficients as a function of the temperature ($T_{e}$) of a Maxwellian EEDF and vibrational temperature of the hydrogen molecules. Calculated using cross sections from Ref. \cite{Janev2003}.}%
 \end{center}
 \end{figure}

The vibrational distribution of low temperature molecular hydrogen plasmas is not in thermal equilibrium (e.g. Refs. \cite{experimental_vibrational_temperature,the_simulation_paper} and references therein). The fraction of molecules at high vibrational levels affects the rates of several plasma processes, e.g. the volume production of negative hydrogen ions through dissociative attachment \cite{Bacal_volume_production}. It has been concluded \cite{Bacal_volume_production,BC_contribution}, that most of the vibrational excitations populating states $\nu \geq 5$ are preceded by electron impact excitation to the $B^1\Sigma^+_u$ and $C^1\Pi_u$ singlet states. Radiative transitions from the given excited electronic states to the ground state populate high vibrational levels, when the vibrational level changes in the electronic transition according to the Franck-Condon principle \cite{HydrogenFranckCondon}. By assuming that transitions between vibrational levels in electron impact excitation to electronic states follow the Franck-Condon factors it can be calculated that 63--67\% of the (de-)excitations emitting in Lyman-band and 46--49\% emitting in Werner-band to ground state vibrational levels $\nu\geq 5$.  Approximately 15\%  of transitions from $B^1\Sigma^+_u$ and $C^1\Pi_u$ states lead to such high vibrational levels (vibrational continuum) that the molecule dissociates instantly \cite{BC_dissociation}. The corresponding dissociation rate does not depend strongly on the electron energy or vibrational temperature of the molecules.

\begin{figure}
 \begin{center}
 \includegraphics[width=0.45\textwidth]{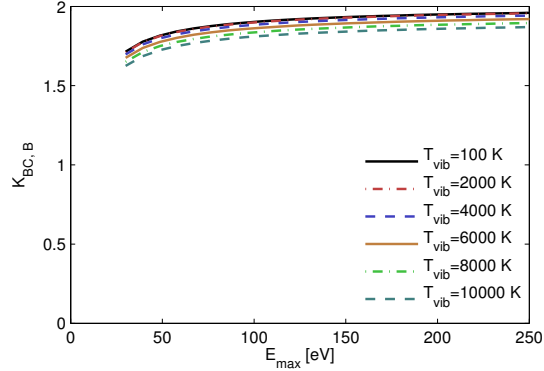}% % Important NOTE: Please make certain your figures do not %include local directory paths. ex. "c:\file\sub\fig1.eps"
 \caption{\label{fig:BC_FD} The ratio $K_{BC,B}$ of $X^1\Sigma^+_g \rightarrow B^1\Sigma^+_u, C^1\Pi_u$ and $X^1\Sigma^+_g \rightarrow B^1\Sigma^+_u$ excitation rate coefficients as a function of the maximum energy ($E_{max}$) of a flat EEDF and vibrational temperature of the hydrogen molecules. Calculated using cross sections from Ref. \cite{Janev2003}.}%
 \end{center}
 \end{figure}

The total excitation rate to $B^1\Sigma^+_u$ and $C^1\Pi_u$ states can be estimated from the measured Lyman-band emission with Eq. \ref{eq:ratio_of_rate_coefficients_simpler} by comparing the total excitation rate coefficients to $B^1\Sigma^+_u$ and $C^1\Pi_u$ states to the excitation rate coefficient to  $B^1\Sigma^+_u$ state. The ratio of the rate coefficients, $K_{BC,B}$, is plotted for different vibrational temperatures and EEDFs in Figs. \ref{fig:BC_MBD} and \ref{fig:BC_FD}.

The rate coefficient ratio is only weakly sensitive to variations of the EEDF in the range of 10-20 eV. This can be explained by the similarity of the functional shapes of the cross sections and threshold energies (Fig. \ref{fig:cross_sections}). For example, using median values of $K_{BC,B}=1.64$ for Maxwellian EEDF or $K_{BC,B}=1.79$ for the flat EEDF cause corresponding uncertainties of \textless$\pm10$\%, if the real electron temperature is in the range of 3~eV--25~eV or the effective $E_{max}$ is in the range of 30~eV--250~eV. On the other hand, using $T_{vib}=6000$~K causes an uncertainty of  \textless$\pm4$\% to $K_{BC,B}$ if the real vibrational temperature is in the range of 100--10000 K. Therefore, the total uncertainty corresponding to the selected median values of $K_{BC,B}$ (in respective ranges of EEDF vibrational temperature variations) is less than 11\%.

\subsection{Maximum production rate of metastable hydrogen molecules}
\label{cap:metastable}
The metastable state  ($c^3\Pi_u$) of the hydrogen molecule has an important role in molecular hydrogen plasmas. Metastable molecules have lower ionization potential and different electronic transition probabilities than ground state molecules. Therefore, the existence of metastable molecules allows ionization and de-excitation by lower energy electrons, which in turn complicates plasma diagnostics based on triplet transitions. Metastable states can also affect the production of negative hydrogen ions \cite{Bacal_volume_production}.

\begin{figure}
 \begin{center}
 \includegraphics[width=0.45\textwidth]{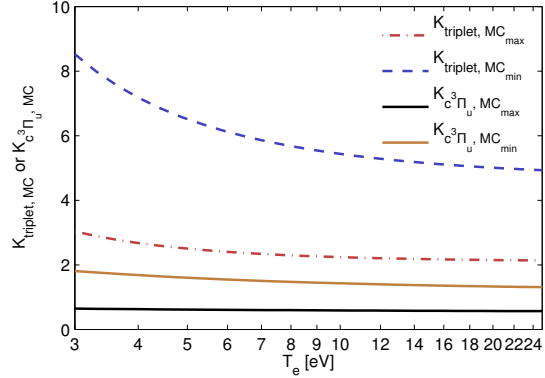}% % Important NOTE: Please make certain your figures do not %include local directory paths. ex. "c:\file\sub\fig1.eps"
 \caption{\label{fig:dissociation_MBD} Ratios of rate coefficients of total triplet ($a^3\Sigma^+_g$, $b^3\Sigma^+_u$, $c^3\Pi_u$, $d^3\Pi_u$, $e^3\Sigma^+_u$) excitation to excitation leading to molecular continuum emission ($K_{triplet,MC}$) and metastable  $c^3\Pi_u^+$ state excitation to excitation leading to molecular continuum emission ($K_{c^3\Pi,MC}$) as a function of temperature ($T_{e}$) of a Maxwellian EEDF. The given range corresponds to the assumption that either all (min) or none (max) excitations to the metastable $c^{3}\Pi_{u}$ state lead to emission within the molecular continuum. Calculated using cross sections from Ref. \cite{Janev2003}.}%
 \end{center}
 \end{figure}

From the diagnostics point-of-view there are significant uncertainties related to the metastable $c^3\Pi_u$ state. Different symmetries of the $c^3\Pi_u$ state ($c^3\Pi_u^+$ and $c^3\Pi_u^-$) have significantly different lifetimes. The lifetime of the $c^{3}\Pi_{u}^{+}$ state is 6.2 ns i.e. only the $c^{3}\Pi_{u}^{-}$ state can be considered metastable \cite{statelifetimes}. There are only few studies about the $c^3\Pi_u^+$ state with a significant discrepancy between calculations and experiments\cite{statelifetimes}. The $c^3\Pi_u$ state is argued to be metastable without any distinction between the parities of the wavefunction ($c^3\Pi_u^-$) in the electron impact cross-section data  (Ref. \cite{Janev2003} and references therein). Because there are no cross section data about electron impact excitation to the $c^3\Pi_u^+$ state, the $c^3\Pi_u$ state is assumed to behave as the metastable $c^3\Pi_u^-$ state. Moreover, experiments and theoretical calculations about electron impact excitation cross sections from the ground state  $X^1\Sigma^+_g$ to $c^3\Pi_u$ state differ by a factor of 2--3 \cite{Janev2003}. There is also significant difference between calculated cross sections e.g. in Refs. \cite{Janev2003} and \cite{electron_impact_triplet_vibrational_dependency}.

From the point-of-view of molecular continuum emission there are three different types of triplet states. The lowest triplet state, $b^3\Sigma_u^+$, is repulsive, and electron impact excitation to the this state leads to non-radiative dissociation.
Triplet states $a^3\Sigma_g^+$, $d^3\Pi_u$ and $e^3\Sigma_u^+$ lead to emission within the molecular continuum. The metastable state $c^3\Pi_u$  can either spontaneously decay, transform to radiative triplet state ($a^3\Sigma_g^+$) by electron impact de-excitation or collisional quenching, the molecule can drift to the wall of the plasma chamber or the molecule can be ionized by electron impact. Therefore, the contribution of the metastable molecules on the molecular continuum emission depends on the plasma parameters and the dimensions of the plasma chamber.

\begin{figure}
 \begin{center}
 \includegraphics[width=0.45\textwidth]{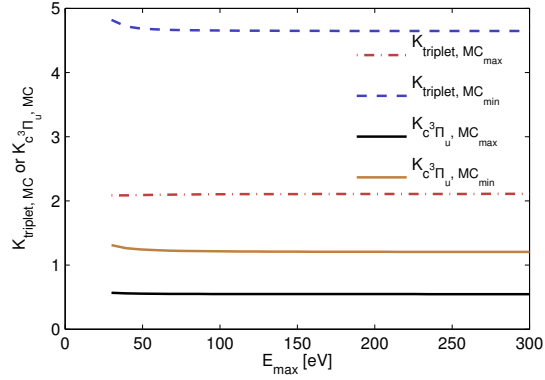}% % Important NOTE: Please make certain your figures do not %include local directory paths. ex. "c:\file\sub\fig1.eps"
 \caption{\label{fig:dissociation_FD}Ratios of rate coefficients of total triplet excitation ($a^3\Sigma^+_g$, $b^3\Sigma^+_u$, $c^3\Pi_u$, $d^3\Pi_u$, $e^3\Sigma^+_u$) to excitation leading to molecular continuum emission ($K_{triplet,MC}$) and metastable  $c^3\Pi_u^+$ state excitation to excitation leading to molecular continuum emission ($K_{c^3\Pi,MC}$) as a function of maximum energy ($E_{max}$) of a flat EEDF. The given range corresponds to the assumption that either all (min) or none (max) excitations to the metastable $c^{3}\Pi_{u}$ state lead to emission within the molecular continuum.  Calculated using cross sections from Ref. \cite{Janev2003}.}%
 \end{center}
 \end{figure} 

The volumetric production rate of metastable molecules  ($c^3\Pi_u^-$) can be estimated from the measured volumetric emission rate of molecular continuum ($a^{3}\Sigma_{g}^{+}\rightarrow b^{3}\Sigma_{u}^{+}$) radiation and the ratio of the rate coefficients of excitation to metastable states and total excitation to triplet states leading to molecular continuum emission. The minimum production rate of metastable molecules can be estimated if most of the excitations to $c^{3}\Pi_{u}$-state are assumed to eventually emit a photon within the molecular continuum (e.g. via collisional quenching or decay to $a^3\Sigma_g^+$). The maximum of metastable state production rate corresponds to the situation in which all metastable states decay without photon emission in the molecular continuum. Because of the large cross section of electron impact excitation from the ground state to the $c^3\Pi_u$ state, there is approximately a factor of two difference between these two extremes.

The uncertainty caused by such approximations (extreme cases) can be reduced by comparing the transition probabilities of different processes affecting the population density of the $c^{3}\Pi_{u}$ state. Typically, the most significant de-excitation processes are electron impact de-excitation, collisional quenching with neutral particles and collision to the wall of the plasma chamber, as discussed in Ref. \cite{Bonnie_1988_mestable_H2_in_multicusp_ion_source}. The radiative lifetime of the $c^{3}\Pi_{u}^-$ state depends on the vibrational level and ranges from \SI{1}{\milli\second} ($\nu$=0) to \SI{10}{\micro\second} ($\nu$=3) \cite{H2_c_lifetime}. These radiative lifetimes correspond to de-excitation transition probabilities of \SI{e3}{1\per\second} ($\nu$=0) and \SI{e5}{1\per\second} ($\nu$=3), respectively.  The collisional quenching rate coefficient of the $c^{3}\Pi_{u}^-$ state in collisions with hydrogen molecules at 300 K is \SI{1.88e-9}{\centi\meter^3\per\second} \cite{H2_c_quenching}. This means that the transition probability of collisional quenching depends on the neutral gas pressure as \SI{4.7e5}{1\per\second\pascal}. The rate coefficient of the most significant electron impact de-excitation process (the superelastic transition $c^{3}\Pi_{u}\rightarrow b^3\Sigma_u^+$) is on the order of \SI{e-7}{\centi\meter^3\per\second} regardless of the electron temperature \cite{transitions_triplet_1}. Thus, collisional de-excitation is significant if the neutral gas pressure is more than \SI{2e-3}{\pascal} and/or the electron density is more than \SI{e10}{\centi\meter^{-3}} for $c^{3}\Pi_{u}$($\nu$=0) or \SI{0.2}{\pascal} and/or \SI{e13}{\centi\meter^{-3}} for $c^{3}\Pi_{u}(v=3)$  state. Furthermore, the $c^{3}\Pi_{u}^-$ state de-excites due to wall collisions \cite{Bonnie_1988_mestable_H2_in_multicusp_ion_source}. This is a significant process in small plasma sources at low neutral gas pressure and electron density, because the average thermal velocity of hydrogen molecules ($T$=300~K, \SI{1.9}{\milli\meter\per\micro\second}) corresponds to a distance of some centimetres during the radiative lifetime of the $c^{3}\Pi_{u}^-$ state.

The ratio of the total rate coefficient of electron impact excitation to metastable state ($c^3\Pi_u$) and the total rate coefficient of electron impact excitation to the molecular continuum radiative states ($a^3\Sigma_g^+$, $d^3\Pi_u$, $e^3\Sigma_u^+$ with/without $c^3\Pi_u$), $K_{c^3\Pi, MC_{min}}$ and $K_{c^3\Pi, MC_{max}}$, is plotted for different EEDFs in Figs. \ref{fig:dissociation_MBD} and \ref{fig:dissociation_FD}. The insensitivity of the rate coefficient ratio to the EEDF can be explained by similar functional shapes and threshold energies of the processes. The contribution of different vibrational levels is not taken into account due to the lack of numerical data \cite{Janev2003,electron_impact_triplet_vibrational_dependency}. However, it can be argued that because increasing vibrational level decreases the threshold energies and increases the cross sections almost identically for electron impact excitation from the ground state to any triplet state \cite{electron_impact_triplet_vibrational_dependency}, the effect of vibrational temperature on $K_{c^3\Pi, MC}$ is almost negligible (of the same order as in Figs.  \ref{fig:BC_MBD} and \ref{fig:BC_FD}).

Using median values of $K_{c^3\Pi, MC_{max}}$=0.60/$K_{c^3\Pi, MC_{min}}$=1.55 (Maxwellian EEDF) and $K_{c^3\Pi, MC_{max}}$=0.56/$K_{c^3\Pi, MC_{min}}$=1.26 (flat EEDF) cause an uncertainty of \textless17\% if the electron temperature is in the range of 3--25~eV or \textless5\% if the  effective $E_{max}$ is in the range of 30--250~eV. The most significant uncertainties (that depend on each other) are, however, caused by the contribution of metastable molecules ($c^3\Pi_u^+$) to the molecular continuum emission and the uncertainty of the cross section data. In the worst case, both of them can cause a factor of two deviation between the diagnostics result and the reality. The most significant factor is the uncertainty of the ratio of excitation cross section to metastable state $X^1\Sigma^+_g \rightarrow c^3\Pi_u$ and total excitation cross section to other triplet states $X^1\Sigma^+_g \rightarrow (a^3\Sigma_g^+$, $b^3\Sigma_u^+$, $d^3\Pi_u$, $e^3\Sigma_u^+$).

\subsection{Molecule dissociation rate via triplet state excitation}
\label{cap:dissociation}

There are several processes which lead to molecule dissociation in hydrogen plasmas e.g. dissociative electron attachment, recombination, ionization, excitation to triplet states, excitation to singlet states on high vibrational level and processes of molecular hydrogen ion \cite{Janev2003}. The significance of each process depends on the plasma parameters. Unfortunately, there are no reliable experimental data for the cross sections of molecule dissociation as discussed in Ref \cite{crosssections2}. It has been argued that electron impact excitation to triplet states is the main dissociative channel for the hydrogen molecule \cite{Celiberto_2001_cross_section_functions,crosssections2}.

The minimum dissociation rate of the molecules by the lowest, repulsive, triplet state $b^{3}\Sigma_{u}^{+}$ can be estimated from the measured molecular continuum emission. In this case, the rate coefficient of electron impact excitation to all triplet states is compared to the rate coefficient of electron impact excitation to the states emitting a photon within the molecular continuum. It can be argued that all of the electron impact excitations to the $c^3\Pi_u$ state lead to the dissociation of the molecule, because both, the radiative and collisional de-excitation of $c^3\Pi_u$ lead mainly to $b^{3}\Sigma_{u}^{+}$ state.

Two extremes of the rate coefficient ratios can be considered by assuming that either all ($K_{triplet, MC_{min}}$)  or none ($K_{triplet, MC_{max}}$) of the metastable states lead to molecular continuum emission. There is approximately a factor of two difference between these extremes. The coefficient ratios $K_{triplet, MC_{min}}$ and $K_{triplet, MC_{max}}$ are plotted for Maxwellian and flat EEDFs in Figs. \ref{fig:dissociation_MBD} and \ref{fig:dissociation_FD}. The error analysis discussed in Section \ref{cap:metastable} for the production rate of metastable $c^3\Pi_u$ is valid also for dissociation rate of hydrogen molecules discussed in this section.  

\subsection{Hot electron density}
\label{cap:hot electron density}

\begin{figure}
 \begin{center}
 \includegraphics[width=0.45\textwidth]{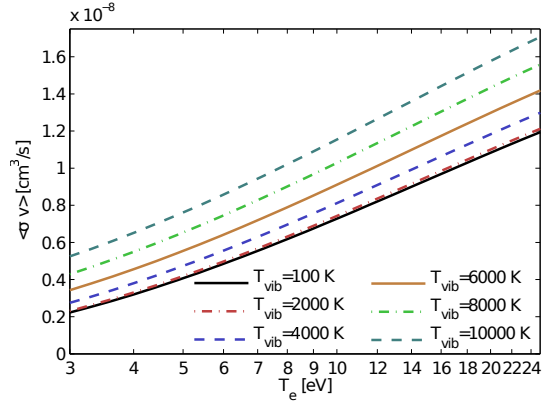}% % Important NOTE: Please make certain your figures do not %include local directory paths. ex. "c:\file\sub\fig1.eps"
 \caption{\label{fig:hot_electrons_MBD}The rate coefficient $\braket{\sigma v}$ of $X^1\Sigma^+_g$ $\rightarrow$ $ B^1\Sigma^+_u$ electron impact excitation for a Maxwellian EEDF as a function of electron temperature ($T_{e}$). Calculated using cross sections from Ref. \cite{Janev2003}.}%
 \end{center}
 \end{figure}
 
\begin{figure}
 \begin{center}
 \includegraphics[width=0.45\textwidth]{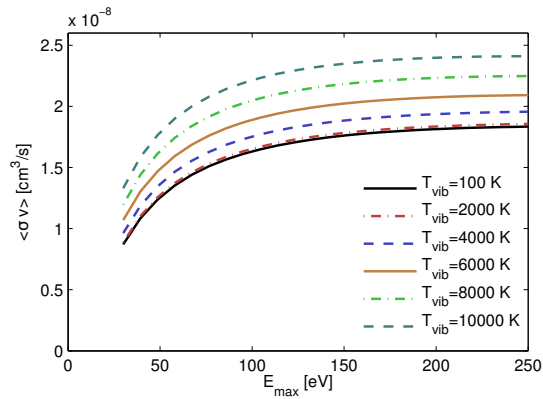}% % Important NOTE: Please make certain your figures do not %include local directory paths. ex. "c:\file\sub\fig1.eps"
 \caption{\label{fig:hot_electrons_FD}The rate coefficient $\braket{\sigma v}$ of $X^1\Sigma^+_g \rightarrow B^1\Sigma^+_u$ electron impact excitation for a flat EEDF as a function of the maximum electron energy ($E_{max}$). Calculated using cross sections from Ref. \cite{Janev2003}.}%
 \end{center}
 \end{figure}

The density of hot electrons ($E_e>9$ eV) can be negligible in comparison to the total electron density of the plasma. However, it  is the only part of the EEDF which produces (positive) ions and electrons from the neutral gas. Therefore, the diagnostic of this population is important especially from the point-of-view of applications, e.g. optimizing the performance of ion sources. The $\braket{\sigma v}$ of $X^1\Sigma^+_g \rightarrow B^1\Sigma^+_u$ electron impact excitation is rather insensitive to the electron energy, which allows estimating the hot electron  density. When the density of neutral molecules $n_n$ is known and the ionization degree is low enough not to affect the neutral density, the hot electron density can estimated with
\begin{equation}
\label{eq:hot_electron_density}
n_e=\frac{R_{Ly-band}}{n_n\braket{v\sigma}},
\end{equation}
where $R_{Ly-band}$ is the measured (volumetric) emission rate of Lyman-band radiation and $\braket{v\sigma}$ is the rate coefficient. The rate coefficient for Maxwellian and flat EEDFs is presented in Figs. \ref{fig:hot_electrons_MBD} and \ref{fig:hot_electrons_FD}. 

It can be concluded from Figs.  \ref{fig:hot_electrons_MBD} and \ref{fig:hot_electrons_FD}, that the uncertainty of $\braket{v\sigma}$ caused by the vibrational temperature varies from 40\% ($T_e$=3~eV) to 20\% ($T_e$=24~eV) in the case of Maxwellian distribution and from 21\% ($E_{max}$=35~eV) to 13\% ($E_{max}$=250~eV) in the case of flat distribution if $T_{vib}=6000$~K is used and the real vibrational temperature is in the range of 100--10000~K. If the electron temperature can be estimated (or measured) with an accuracy of $\pm20$\%, it causes an uncertainty of \textless$\pm$15\% to $\braket{v\sigma}$. Using the median value of $\braket{v\sigma}=$\SI{1.6e-8}{\centi\meter^{3}\per\second} causes an uncertainty of \textless$\pm$33\% if the effective $E_{max}$ is in the range of 30--250~eV. Thus, the total uncertainty of $\braket{v\sigma}$ is \textless$\pm$43\% for Maxwellian EEDF and \textless$\pm$39\% for the flat EEDF in the described situations.

\subsection{Error analysis}
\label{cap:error_analysis}
The presented methods to analyse reaction rates in molecular hydrogen plasma are based on absolute measurements of the volumetric emission rates of specific emission bands. There are several sources of error affecting the estimated absolute reaction rates:
\begin{enumerate}[i)]
\item\label{item:error_measurement} measurement technique
\item\label{item:error_band}  overlapping emission bands
\item\label{item:error_cover}  coverage of emission band by the measurement
\item\label{item:error_cascade} cascade from upper states
\item\label{item:error_EEDF}  deviation of the assumed EEDF from the real one
\item\label{item:error_vib}  estimation of the vibrational distribution
\item\label{item:error_other_processes}  effect of other plasma processes e.g. collisions of metastable molecules
\item\label{item:error_cross_section}  uncertainty of the cross sections
\item\label{item:error_assumptions} validity of the assumptions listed in section \ref{sec:population_density}
\end{enumerate}

In an earlier paper \cite{komppula_NIBS_2012}, it was estimated that the uncertainty of the measured volumetric emission rates of specific emission band in specific wavelength range is less than 18\% (the experimental setup is described briefly in section \ref{sec:example_analysis}). This error covers both, the error of the measurement itself (item \ref{item:error_measurement}) and the overlap of the emission bands (item \ref{item:error_band}) in the selected wavelength ranges of 145--170~nm for Lyman-band and 170--240~nm for molecular continuum. 

The measured ranges of Lyman-band and molecular continuum emission do not cover the entire emission band. The coverage can be estimated by studying synthetic emission spectra. The synthetic spectrum can be calculated, if vibrational the distribution of the upper state and the Franck-Condon factors governing the change of vibrational level in electronic transitions are known. The vibrational distribution of the $B^1\Sigma^+_u$ state can be estimated by using distributions described in Ref. \cite{LymanWernerExperiment} or by assuming that the vibrational temperature of ground state hydrogen molecules is between 0--10000~K and that vibrational excitation in the electron impact excitations to the $B^{1}\Sigma_{u}$ electronic state follows the Franck-Condon factors (first order approximation). Using the afore-mentioned vibrational distributions at $B^1\Sigma^+_u$ state and Franck-Condon factors from Ref. \cite{HydrogenFranckCondon} yields that the range of 145--170~nm covers 38--43\% of the total Lyman-band emission. Using the synthetic spectrum of molecular continuum from Ref. \cite{FantzMolC}, yields that the range 170--240~nm covers 59--62\% of the total molecular continuum emission if the vibrational temperature of the neutral gas is in the range of 0--9000~K. If median values (40\% and 61\% respectively) are used, the variation of the correction factors correspond to uncertainties of 5\% and 3\% for Lyman-band and molecular continuum emissions, respectively.

Recent cross section data of electron impact excitation to $EF^1\Sigma_g$ state \cite{Liu_2003_Electron_impact_excitation_of_H2_EF} imply that 9--15\% of the total Lyman-band emission is caused by cascade from the $EF^1\Sigma_g$ state in the case of the studied EEDFs. Synthetic spectra ($T_{vib}=$100--10000~K) imply, that the measured signal (145--170~nm)  includes  31--34\% of the total cascade effect. Thus, 86--93\% of the measured signal is caused by direct excitation to $B^1\Sigma^+_u$ state. Using a median value (89.5\%) for the correction includes a relative uncertainty of 4\% (estimated similar to the previous paragraph). Cascading from upper states affects also the molecular continuum. Without accurate cross sections the correction due to cascade contribution (from states which are not taken into account in this study) to molecular continuum emission can not be estimated accurately. However, it can be argued that the cascade effect is small (i.e. \textless$10\%$), because the cross sections of electron impact decrease significantly as a function of the main quantum number.

The total uncertainty caused by items \ref{item:error_measurement}--\ref{item:error_cascade} could be reduced at least by a factor of two with improved calibration of the photodiode and more detailed analysis of overlapping emission bands, filter transmission and stray emission.

The uncertainty caused by estimating the EEDF and vibrational temperature (items \ref{item:error_EEDF} and \ref{item:error_vib}) are discussed in Sections \ref{cap:ionization}--\ref{cap:hot electron density}. The uncertainty is 11\%--25\% in the case of listed plasma processes and 26--43\% in the case of the hot electron density. These values correspond to situations in which the vibrational distribution of the plasma is unknown and only rough approximation of the EEDF is available. Therefore, detailed knowledge on the EEDF and vibrational distribution would reduce these uncertainties significantly. 

The lack of experimental data on relevant cross sections is problematic \cite{crosssections2,Janev2003}. The calculations of the cross sections have become more accurate, but experimental studies include uncertainties of several tens of percents especially at low electron energies \cite{crosssections2}. Moreover, there is a variation from several tens of percent up to a factor of 2--3 between calculations especially in the case of excitation to triplet states \cite{Janev2003,electron_impact_triplet_vibrational_dependency}. The presented diagnostics of plasma processes is mostly sensitive to relative error between the cross sections due to the comparison of rate coefficients. The absolute error of the cross sections does not have a significant effect on the diagnostics results if the main contribution is systematic error, which is similar for each cross section. The uncertainty caused by the cross section data can not be estimated accurately based on the available data, but it can be argued that it is probably comparable to or larger than the total uncertainty caused by items \ref{item:error_measurement}--\ref{item:error_vib}.

Deviations from the assumptions made in Section \ref{sec:population_density} can be considered a minor source of error. This is because typical ion source plasmas are within the range of these assumptions by a great margin (orders of magnitude) \cite{plasma_sources,Bacal_volume_production}.

\section{An example of plasma analysis}
\label{sec:example_analysis}

\begin{figure*}
 \begin{center}
 \includegraphics[width=0.80\textwidth]{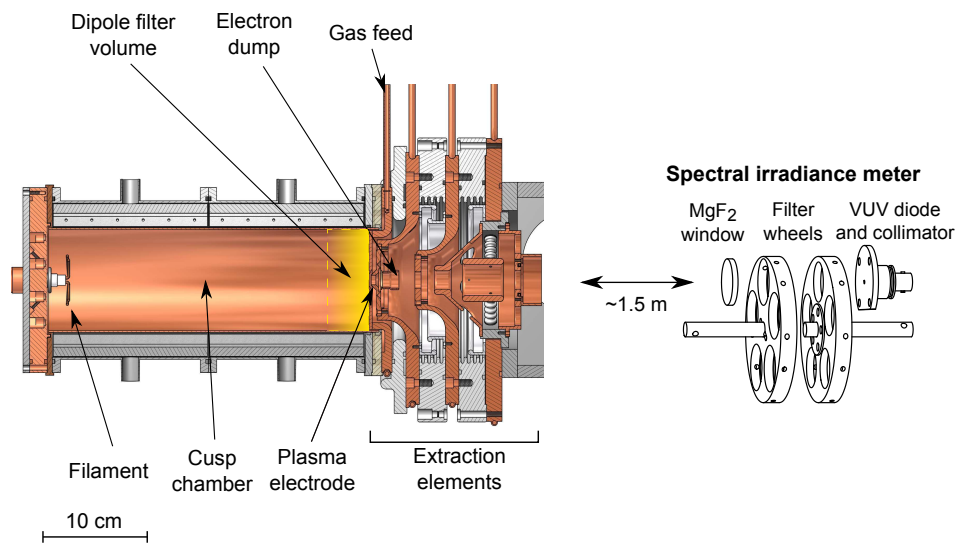}% % Important NOTE: Please make certain your figures do not %include local directory paths. ex. "c:\file\sub\fig1.eps"
 \caption{\label{fig:liisa} LIISA H$^-$ ion source and the irradiance meter used in Ref. \cite{komppula_NIBS_2012}}%
 \end{center}
 \end{figure*}

Absolute and relative VUV emission have been measured earlier from the filament driven arc discharge negative ion source, LIISA \cite{komppula_NIBS_2012} (Fig. \ref{fig:liisa}). The plasma chamber (diameter 9 cm, length 31 cm) is made of copper, but it is coated with tantalum due to the evaporation of the filament. The arc discharge power was varied in the range of 140-840~W by adjusting both, the arc current between 4-12 A and the arc voltage between 35-70 V. The presented values were measured at the plasma chamber pressure of 0.35 Pa, which corresponds 5 sccm feed rate of hydrogen gas\footnote{The optimum plasma chamber pressure is presented incorrectly in Ref. \cite{komppula_NIBS_2012}. The correct plasma chamber pressure is 0.35 Pa (measured with Pirani gauge including up to factor 2 uncertainty \cite{TPR010}) instead of 0.17 Pa . The gas feed rate of 15 sccm was estimated in Ref. \cite{komppula_NIBS_2012}. The value 5 sccm is experimentally verified.}. The relative VUV-emission was not observed to depend significantly on the pressure in the range between \SI[per-mode=fraction]{1.2e-3}{\milli\bar} and \SI[per-mode=fraction]{2.0e-2}{\milli\bar} \cite{komppula_NIBS_2012}. The total electron density of a similar ion source \cite{TRIUMFCharac} has been measured to be always below $10^{12}$~1/cm$^3$. Thus, the requirements for the diagnostics described in the Section \ref{sec:population_density} are valid in the studied ion source.

The measurements were performed with a VUV-irradiance meter, which consists of a factory calibrated photodiode (IRD-inc SXUV20BNC) and optical bandpass filters of specific wavelength ranges -- Lyman-band (161 nm, FWHM 20 nm) and molecular continuum (180 nm, FWHM 40 nm). The device was looking into the plasma along an axial line-of-sight through the extraction aperture (diameter 9~mm) of the ion source from a distance of approximately 1.5 m. The photodiode (an effective area collimated to 3.14~mm$^2$) observed a plasma volume of approximately 20 cm$^3$ out of the 2340 cm$^3$ total volume of the plasma chamber. The transmittances of the filters were measured by using a spectrometer, which consists of a monochromator (McPherson Model 234/302) and a photomultiplier (ET Enterprices 9406B) and utilizing the ion source as a light source. The background signals caused by the stray transmittance and overlapping of adjacent emission bands were taken into account, and the corresponding uncertainty due to such corrections was included to the total uncertainty. Thus, the presented values (Table \ref{tab:old_results}) correspond to Lyman-band emission and molecular continuum emission in the wavelength ranges of 145-170~nm and 170-240~nm, respectively. The average wavelengths in Table \ref{tab:old_results} correspond to energy weighted average wavelengths of the (corrected) emission in the given wavelength range.  

The confinement of primary electrons (emitted by the filament) in the multicusp magnetic field of filament arc discharges is known to be effective (e.g. Refs. \cite{the_book,multicusp_confinement_1}). The same conclusion has been derived from the VUV-emission data obtained with the LIISA ion source. It has been observed that the arc power is efficiently absorbed in inelastic collisions between primary electrons and neutrals \cite{komppula_NIBS_2012}. Furthermore, it has been observed that Lyman band radiation is linearly proportional to the arc discharge power, while the molecular continuum radiation is mainly proportional to the arc discharge current in the entire parameter range. Therefore, it is convenient to transform the measured irradiance values to average volumetric emission rates in the line-of-sight volume and normalize the results to the arc power (Lyman-band) or the arc current (molecular continuum) as presented in Table \ref{tab:old_results}.

The EEDF of the hot electron component in filament arc discharges is best described by the flat distribution discussed in Section \ref{sec:diagnostics}. Such distribution has been observed experimentally and deduced numerically (e.g. Refs. \cite{flat_EEDF_1,flat_EEDF_2,flat_EEDF_3}). Furthermore, the electron energies can not exceed the potential difference of the filament and the plasma, i.e. the Maxwellian distribution would be a non-physical representation of the high energy tail.

 \begin{table*}
 \caption{\label{tab:old_results}Absolute VUV-emission of the filament arc discharge \cite{komppula_NIBS_2012}. Relative error for emission power is $\pm18\%$.}  
 \begin{indented}
 \item[]\begin{tabular}{@{} l r  c  c  l}
 \br
 Emission band & \centre{1}{$\braket{\lambda}$} & Emission power & Photon emission  \\
\mr
Lyman-band/$P_{arc}$ & 160~nm& \SI[per-mode=fraction]{40}{\mW\per\kW~\cm^3} & \SI[per-mode=fraction]{3.3e16}{1\per\kW~\s~\cm^3}\\
Mol. cont/$I_{arc}$ & 195~nm& \SI[per-mode=fraction]{0.55}{\mW\per\A~\cm^3} & \SI[per-mode=fraction]{5.4e14}{1\per\A~\s~\cm^3}\\
\br
\end{tabular}
\end{indented}
 \end{table*}

 \begin{table*}
 \caption{\label{tab:reaction_rates} Calculated reaction rates of different processes and hot electron densities. Variation of the rates corresponds to 35 - 70 eV maximum energies of the flat EEDF. The rates have been normalized with the arc power or arc current depending on their response to these parameters.} 
 \begin{indented}
 \item[]\begin{tabular}{@{} l  c  c  l}
 \br
 &\centre{2}{Vibrational temperature}&\\
 \ns\ns
 &\crule{2}&\\
 Process &  $100$~K& $10000$~K & \\
\mr
Ionization & $8-15$ & $7-12$ & $\times$\SI[per-mode=fraction]{e16}{1\per\kW~\s~\cm^3}\\

Excitation ($B^1\Sigma^+_u$, $C^1\Pi_u$) & $13-14$ & $12-13$ & $\times$\SI[per-mode=fraction]{e16}{1\per\kW~\s~\cm^3}\\

Excitation ($\nu\geq5$) ($B^1\Sigma^+_u$, $C^1\Pi_u$) & $7.4-7.8$ & $7.0-7.4$ & $\times$\SI[per-mode=fraction]{e16}{1\per\kW~\s~\cm^3}\\

Dissociation ($B^1\Sigma^+_u$, $C^1\Pi_u$) & $1.9-2.0$ & $1.8-1.9$ & $\times$\SI[per-mode=fraction]{e16}{1\per\kW~\s~\cm^3}\\

Dissociation (via $b^3\Sigma^+_u$), min & $1.8$ &  & $\times$\SI[per-mode=fraction]{e15}{1\per\A~\s~\cm^3}\\

Dissociation (via $b^3\Sigma^+_u$), max  & $4.2$ & & $\times$\SI[per-mode=fraction]{e15}{1\per\A~\s~\cm^3}\\

$c^3\Pi_u$ states, min & $0.5$ & & $\times$\SI[per-mode=fraction]{e15}{1\per\A~\s~\cm^3}\\

$c^3\Pi_u$ states, max & $1.1$ & & $\times$\SI[per-mode=fraction]{e15}{1\per\A~\s~\cm^3}\\

Hot electron density & $5.7-8.5$ & $4.1-5.7$ & $\times$\SI[per-mode=fraction]{e10}{1\per\kW~\cm^3}\\
\br
\end{tabular}
\end{indented}
 \end{table*}

The estimated volumetric rates and hot electron densities in the line-of-sight volume are presented in Table \ref{tab:reaction_rates}. The total emission rate of Lyman-band (direct excitation to $B^1\Sigma^+_u$ state) and molecular continuum are calculated by using the equation
\begin{equation}
	R_{tot}=\frac{R_{meas}}{k_{cov}}k_{cas},
\end{equation}
where $R_{meas}$ is the measured photon emission rate (Table \ref{tab:old_results}), $k_{cov}$ is a correction factor taking into account the coverage of the measurement compared to the total emission and $k_{cas}$ is a correction factor taking into account the cascade effect. Values of 0.4 (Lyman band) and 0.6 (molecular continuum) have been used for $k_{cov}$ and 0.895 (Lyman band) and 1 (molecular continuum) for $k_{cas}$ as discussed in Section \ref{cap:error_analysis}. The volumetric rates were calculated from equation \ref{eq:ratio_of_rate_coefficients_simpler} by using rate coefficient ratios from Figs. \ref{fig:ionization_FD}, \ref{fig:BC_FD} and \ref{fig:dissociation_FD}. The dissociation rate and vibrational excitation rate to $X^1\Sigma^+_g(v\geq 5)$ via electronic excitation to $B^1\Sigma^+_u$ and $C^1\Pi_u$ states are calculated by using the fractions between the measured and total rates discussed in Section \ref{cap:vibrational_excitation}. The hot electron densities are estimated with Eq. \ref{eq:hot_electron_density} by using the rate coefficient from Fig. \ref{fig:hot_electrons_FD} and neutral gas density corresponding to 0.35~Pa pressure at room temperature. Variation of the values in each cell corresponds to the variation of applied arc voltages ($E_{max}$). Vibrational temperatures 100~K and 10000~K have been used to demonstrate the effect of the vibrational temperature.

The maximum density of metastable molecules can be calculated from the estimated production rate and average lifetime of the metastable states. The plasma parameters (density and neutral gas pressure) are in the range where, in principle, all of the de-excitation processes listed in Section \ref{cap:metastable} affect the effective lifetime of the $c^3\Pi_u$ state. However, the maximum effective lifetime of the $c^3\Pi_u$ state can be estimated to be \SI[per-mode=symbol]{26}{\micro\second}, which corresponds to the time-of-flight of thermal hydrogen molecule across the radius of the plasma chamber (5~cm). Hence, a typical maximum production rate of \SI[per-mode=symbol]{1.1e+16}{1\per\second\centi\meter^3} with 10 A arc current corresponds to a maximum density of \SI[per-mode=symbol]{4e+11}{1\per\centi\meter^3} molecules excited to the $c^3\Pi_u$ state. This in turn corresponds to approximately 0.4\% of the neutral gas density at 0.35 Pa at room temperature.

 \begin{table*}
 \caption{\label{tab:number_of_reactions_per_molecule} The average number of a single molecule to undergo a certain process. The numbers have been normalized with the arc power or arc current. The calculated numbers assume a homogeneous and isotropic plasma emission distribution occupying the whole chamber. This assumption overestimates the values by not more than 50\% as explained in Ref. \cite{komppula_NIBS_2012}.} 
 \begin{indented}
 \item[]\begin{tabular}{@{} l  c  c  l}
 \br
 &\centre{2}{Vibrational temperature}&\\
 \ns\ns
 &\crule{2}&\\
 Process &  $100$~K& $10000$~K & \\
\mr
Ionization & $73-141$ & $61-117$ & \SI[per-mode=fraction]{}{1\per\kW}\\

Excitation ($B^1\Sigma^+_u$, $C^1\Pi_u$) & $121-129$ & $115-123$ & \SI[per-mode=fraction]{}{1\per\kW}\\

Excitation ($\nu\geq5$) ($B^1\Sigma^+_u$, $C^1\Pi_u$) & $70-74$ & $67-70$ & \SI[per-mode=fraction]{}{1\per\kW}\\

Dissociation ($B^1\Sigma^+_u$, $C^1\Pi_u$) & $18-19$ & $17-18$ & \SI[per-mode=fraction]{}{1\per\kW}\\

Dissociation (via $b^3\Sigma^+_u$), min & $1.7$ & & \SI[per-mode=fraction]{}{1\per\A}  \\

Dissociation (via $b^3\Sigma^+_u$), max & $3.9$ & & \SI[per-mode=fraction]{}{1\per\A}  \\

$c^3\Pi_u$ states, min & $0.5$ & & \SI[per-mode=fraction]{}{1\per\A}  \\

$c^3\Pi_u$ states, max & $1.0$ & & \SI[per-mode=fraction]{}{1\per\A}  \\
\br
\end{tabular}
\end{indented}
 \end{table*} 

The total reaction rates can be compared to the neutral gas density, the gas feed rate and the number of electrons emitted by the filament. This allows estimating the average number of different reactions experienced by a single molecule during the time it spends in the plasma chamber (Table \ref{tab:number_of_reactions_per_molecule}) and the average number of different types of inelastic collisions caused by a single electron emitted by the filament (Table \ref{tab:number_of_reactions_per_electron}). The total reaction rates can be estimated by assuming homogeneous and isotropic plasma emission profile. This overestimates the obtained values by less than 50\% \cite{komppula_NIBS_2012}.

The gas feed rate of 5 sccm corresponds to \SI[per-mode=fraction]{2.1e18}{molecules\per\s}. Comparing this number to the total number of molecules in the chamber (0.35~Pa of ideal gas, N=\SI[per-mode=fraction]{2.6e17}{molecules}) yields an average passage time of \SI[per-mode=fraction]{83}{\ms} through the plasma chamber. This means that each molecule undergoes on average several tens of ionization and singlet state excitation reactions and dissociates several times (Table \ref{tab:number_of_reactions_per_molecule}). 

It is possible to estimate the energy efficiency of the plasma source by comparing the total reaction rates to the heating power and the number of electrons emitted by the filament. This is presented in Table \ref{tab:number_of_reactions_per_electron}. Each kilowatt of discharge power produces approximately 1.3--\SI[per-mode=symbol]{3.0e20}{} of electrons and ions per second via ionization corresponding to 20-40 Amperes of electrical current. Approximately the same number of neutral molecules are excited to singlet states  per kilowatt of discharge power. The molecular continuum emission is proportional to the arc current, and therefore, it is reasonable to compare the molecule dissociation rate and the production rate of metastable $c^3\Pi_u^+$ states to the number of electrons emitted by the filament. It turns out that every electron emitted by the filament dissociates on average 0.6-1.3 molecules via triplet states and excites 0.15-0.4 molecules to the metastable state. The dissociation rates via triplet and singlet states are comparable under typical ion source parameters (arc voltage 70 V, arc current 12 A), i.e. molecules dissociate only 1.5--3 times more frequently via triplet states than singlet states.  The results are consistent with the explanation given in Ref. \cite{komppula_NIBS_2012}. There, the conclusion was that during the thermalization process each electron emitted by the filament passes through the optimum energy range for triplet excitation, which explains the molecular continuum radiation being proportional to the arc current.

 \begin{table*}
 \caption{\label{tab:number_of_reactions_per_electron} The average number of ionization, excitation, dissociation and production of metastable states per kilowatt of arc discharge power or electrons emitted by the filament. The calculated numbers assume a homogeneous and isotropic plasma emission distribution occupying the whole chamber. This assumption overestimates the values not more than 50\% as explained in Ref. \cite{komppula_NIBS_2012}.} 
 \begin{indented}
 \item[]\begin{tabular}{@{} l  c  c  l}
 \br
 &\centre{2}{Vibrational temperature}&\\
 \ns\ns
 &\crule{2}&\\
 Process &  $100$~K& $10000$~K & \\
\mr
Ionization & $15-30$ & $13-24$ & $\times$\SI[per-mode=fraction]{e19}{1\per\kW\second}\\

Excitation ($B^1\Sigma^+_u$, $C^1\Pi_u$) &  $25-27$ &  $24-26$ & $\times$\SI[per-mode=fraction]{e19}{1\per\kW\second}\\

Excitation ($\nu\geq5$) ($B^1\Sigma^+_u$, $C^1\Pi_u$) &  $15$ &  $14-15$ & $\times$\SI[per-mode=fraction]{e19}{1\per\kW\second}\\
Dissociation ($B^1\Sigma^+_u$, $C^1\Pi_u$) &  $3.8-4.0$ &  $3.6-3.8$ & $\times$\SI[per-mode=fraction]{e19}{1\per\kW\second}\\

Dissociation (via $b^3\Sigma^+_u$), min &  $0.6$ &  & events/$e_{arc}$ \\

Dissociation (via $b^3\Sigma^+_u$), max &  $1.3$ &  & events/$e_{arc}$\\

$c^3\Pi_u$ states, min &  $0.15$ &  & events/$e_{arc}$ \\

$c^3\Pi_u$ states, max &  $0.35$ &  & events/$e_{arc}$\\
\br
\end{tabular}
\end{indented}
 \end{table*}

\section{Discussion}
\label{sec:discussion}

The principles of robust and straightforward diagnostic of electron impact processes in low temperature molecular hydrogen plasmas have been presented in this study. The method could contribute significantly to the development of plasma and ion sources since the effects of mechanical modifications, for example, could be connected directly to the changes of reaction rates in the plasma. Similar information could be obtained by measuring the plasma parameters (neutral gas density, electron density and EEDF) and applying collision radiative models.

\begin{figure}
 \begin{center}
 \includegraphics[width=0.45\textwidth]{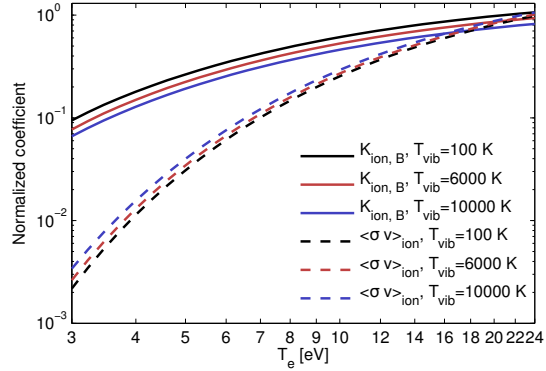}% % Important NOTE: Please make certain your figures do not %include local directory paths. ex. "c:\file\sub\fig1.eps"
 \caption{\label{fig:uncertainty_comparison} Normalized ionization rate coefficient $\Braket{\sigma v}_{ion}$ and ratio of rate coefficients ($K_{ion,B\Sigma}$) as a function of electron temperature in different vibrational temperatures.}%
 \end{center}
 \end{figure}

The presented methods include significantly smaller uncertainty than calculating the reaction rates from the measured plasma parameters. This is because applying the presented method does not require any information about the electron and neutral densities and because the rate coefficient ratio $K_{2,1}$ (Eq. \ref{eq:ratio_of_rate_coefficients_simpler}) is less sensitive to the electron temperature than the rate coefficient $\Braket{\sigma v}$ itself. The most significant difference in the uncertainties of different methods is caused by the electron temperature dependence of $K_{2,1}$ and $\Braket{\sigma v}$. This is demonstrated in Fig. \ref{fig:uncertainty_comparison} showing the normalized ratio of ionization and $B^{1}\Sigma_{u}^{+}$ excitation rate coefficients $K_{ion,B}$ and normalized ionization rate coefficient $\Braket{\sigma v}_{ion}$ as a function of the electron temperature. Although $K_{ion,B}$ is the most sensitive $K_{2,1}$ coefficient (discussed in this study) to the variation of the electron temperature, it changes only by an order of magnitude in the range of 3~eV\textless$T_e$\textless25~eV, while $\Braket{\sigma v}_{ion}$ varies almost three orders of magnitude. For the other processes studied in this paper the difference is even more significant. The coefficient $K_{ion,B}$ is more sensitive to the variation of the vibrational temperature than $\Braket{\sigma v}_{ion}$. However, for other processes, $K_{2,1}$ is less sensitive to vibrational temperature than $\Braket{\sigma v}$ due to similar dependence of the rate coefficients on the vibrational temperature. 

Based on the error analysis presented in Section 3.6. it is justified to claim that the main sources of uncertainty for determining the ionization rate and total excitation rate to $B^1\Sigma^+_u$ and $C^1\Pi_u$ states are uncertainties of measurement (relative uncertainty $\pm18$\%), the coverage of the measured emission ($\pm5$\%), cascade effects ($\pm4$\%) and determining the coefficients $K_{ion,B}$ and $K_{BC,B}$ (\textless$\pm25$\% within the limits of EEDF and vibrational temperature variations discussed in Sections 3.1 and 2.1). Applying the general law of error propagation for the independent error sources, it can be argued that with the presented methods it is possible to measure the ionization rate and total excitation rate $B^1\Sigma^+_u$ and $C^1\Pi_u$ with an accuracy of 32\% or better. This number does not take into account the uncertainty of the cross section data. The both effect of metastable states and discrepancies in cross section data (e.g. in Ref.s [12] and [39]) do not allow making a general statement about the uncertainty of diagnostics results based on molecular continuum emission. However, the EEDF, the vibrational distribution and the density of metastable molecules are often characteristic properties of the given plasma source. Therefore, the listed uncertainties mostly affect the systematic error of the VUV diagnostic. It is possible to measure relative changes with significantly higher accuracy in comparison to absolute values, which is often more important for practical development of plasma sources.

Several simulation codes have been developed for molecular hydrogen plasmas during the last decades, e.g. Ref. \cite{the_simulation_paper} and references therein. The presented diagnostics of the reaction rates could be a powerful tool to benchmark the simulation codes, as introduced in Ref. \cite{nonMaxwelliansimulation1}. The VUV-emission is a consequence of the most significant plasma processes (with the largest cross sections), which have been taken account by most of the codes. Therefore it should be possible to extract the volumetric VUV emission rates as simulation outputs. Comparison of the simulations and the VUV-spectroscopy could validate input parameters of the simulations, e.g. the functional shape of the hot part of the EEDF and/or the vibrational distribution. The presented diagnostics could allow developing new type of plasma simulations in which the measured reaction rates of hot electron processes (ionization, dissociation and vibrational excitation rates) could be used as a parameter and the simulation could focus on cold electron processes such as plasma diffusion, recombination and dissociative electron attachment.

The presented methods have been applied to analyze a filament arc discharge. Although the plasma parameters depend on mechanical design of the plasma source, some results can be generalized. It is well known, that the confinement of primary electrons emitted by the filament is good in multicusp arc discharges and their energy dissipates mainly in inelastic collisions with neutrals. Therefore, the reaction rates normalized with respect to the arc power and the arc current (Table \ref{tab:number_of_reactions_per_electron}) should be general properties of this type of plasma sources. 

Excitation rate to vibrational levels ($\nu\geq 5$) via electron impact excitation to $B^1\Sigma^+_u$ and $C^1\Pi_u$ state has been estimated earlier by Graham \cite{Graham_1984_VUV_measurement} to be on the order of \SI[per-mode=fraction]{e15}{\cm^{-3}\second^{-1}\kilo\watt^{-1}}. This is approximately an order of magnitude lower than the estimated value given in this paper. The deviation is probably caused by the difference in the confinement of the primary electrons since the apparatus used in Ref. \cite{Graham_1984_VUV_measurement} does not contain a multicusp magnetic field. The density of metastable $c^3\Pi_u^-$ states in a multicusp ion source has been measured earlier to be on the order of \SI[per-mode=fraction]{e10}{\cm^{-3}\kilo\watt^{-1}} \cite{Bonnie_1988_mestable_H2_in_multicusp_ion_source}. This is approximately two orders of magnitude lower than the upper limit estimated in this study. The difference is probably caused by uncertainties related to the estimate presented in this study i.e. a) the applied cross section is the total cross section to $c^3\Pi_u$ states, b) the contribution of $c^3\Pi_u$ state to the molecular continuum emission c) the estimated maximum effective lifetime of the $c^3\Pi_u^-$ state and d) the assumption of homogeneous and isotropic plasma emission profile. Thus, the estimated maximum $c^3\Pi_u^-$ density presented in this study is consistent with the measured value given in Ref. \cite{Bonnie_1988_mestable_H2_in_multicusp_ion_source}.

The benefits of VUV-spectroscopy arise from the quantum mechanical properties of the hydrogen molecule. Processes, like excitation to  triplet states leading to molecule dissociation and excitation to singlet states leading to production of high vibrational levels on the ground state, are directly related to the properties of the hydrogen molecule orbitals and Franck-Condon principle. Although those properties are not valid for other elements, there are often interesting (chemical) phenomena related to electronic excitations and forbidden transitions, e.g. the production of the metastable helium \cite{metastable_helium}. Therefore, it should be possible to utilize the same principle, namely the comparison of the rate coefficients with similar functional shapes, for other plasmas as well. 

\section*{Acknowledgements}
This work has been supported by the EU 7th framework programme 'Integrating Activities -- Transnational Access', project number: 262010
(ENSAR) and by the Academy of Finland under the Finnish Centre of Excellence Programme 2012-
2017 (Nuclear and Accelerator Based Physics Research at JYFL)

\section*{References}

\bibliography{hydrogenplasmarefs}

\begin{thebibliography}{10}

\bibitem{Bacal_volume_production}
M.~Bacal, A.~Hatayama, and J.~Peters.
\newblock {\em IEEE Trans. Plasma Sci}, 33(6):1845--1871, 2005.

\bibitem{H_2_plus_source_1}
K.~W. Ehlers and K.~N. Leung.
\newblock {\em Rev. Sci. Instrum}, 54(6):677--680, 1983.

\bibitem{H_2_plus_source_2}
M.~D. Williams, K.~N. Leung, G.~M. Brennen, and D.~R. Burns.
\newblock {\em Rev. Sci. Instrum}, 61(1):475--477, 1990.

\bibitem{H_2_plus_source_3}
J.~R. Alonso, W.~A. Barletta, M.~H. Toups, J.~Conrad, Y.~Liu, M.~E. Bannister,
  C.~C. Havener, and R.~Vane.
\newblock {\em Rev. Sci. Instrum}, 85(2):02A509, 2014.

\bibitem{H_2_fusion}
R.E.H. Clark and D.~Reiter.
\newblock {\em Nuclear Fusion Research: Understanding Plasma-Surface
  Interactions}.
\newblock Springer Series in Chemical Physics. Springer, 2006.

\bibitem{H_2_astrophysics}
G.~Shaw, G.~J. Ferland, N.~P. Abel, P.~C. Stancil, and P.~A.~M. van Hoof.
\newblock {\em Astrophys. J.}, 624(2):794, 2005.

\bibitem{allHydrogenEmissionSpectroscopy}
M.~Capitelli et~al.
\newblock {\em J Phys. B - At. Mol. Opt}, 43(14):144025, 2010.

\bibitem{molecular_visible_diagnostics1}
K.~Sawada and T.~Fujimoto.
\newblock {\em J. Appl. Phys.}, 78(5):2913--2924, 1995.

\bibitem{Graham_1984_VUV_measurement}
W.~G. Graham.
\newblock {\em J. Phys. D: Appl. Phys.}, 17(11):2225, 1984.

\bibitem{Lavrov_1999_continuum}
B.~P. Lavrov, A.~S. Melnikov, M.~K\"aning, and J.~R\"opcke.
\newblock {\em Phys. Rev. E}, 59:3526--3543, 1999.

\bibitem{komppula_NIBS_2012}
J.~Komppula, O.~Tarvainen, S.~L{\"a}tti, T.~Kalvas, H.~Koivisto, V.~Toivanen,
  and P.~Myllyperki{\"o}.
\newblock {\em AIP Conf. Proc.}, 1515(1):66--73, 2013.

\bibitem{Janev2003}
R.~K. Janev, D.~Reiter, and U.~Samm.
\newblock {\em {C}ollision processes in low-temperature hydrogen plasmas},
  volume JUEL-4105 of {\em Berichte des Forschungszentrums J\"{u}lich}.
\newblock Forschungszentrum, Zentralbibliothek, J\"{u}lich, 2003.

\bibitem{H_Franck-Condon_factors}
U.~Fantz and D.~W\"underlich.
\newblock {\em Atom. data nucl. data}, 92(6):853--973, 2006.

\bibitem{H2_c_lifetime_calculation}
E.~E. LaFleur and L.~C. Chiu.
\newblock {\em J. Chem. Phys.}, 84(4):2150--2157, 1986.

\bibitem{metastable_calculation_Freis}
R.~P. Freis and J.~R. Hiskes.
\newblock {\em Phys. Rev. A}, 2:573--580, 1970.

\bibitem{statelifetimes}
S.A. Astashkevich and B.P. Lavrov.
\newblock {\em Opt. Spectrosc.}, 92(6):818--850, 2002.

\bibitem{transitions_triplet_1}
C.~S. Sartori, F.~J. da~Paix\~ao, and M.~A.~P. Lima.
\newblock {\em Phys. Rev. A}, 58:2857--2863, Oct 1998.

\bibitem{transitions_triplet_2}
A.~Laricchiuta, R.~Celiberto, and R.~K. Janev.
\newblock {\em Phys. Rev. A}, 69:022706, Feb 2004.

\bibitem{H2_c_quenching}
A.~B. Wedding and A.~V. Phelps.
\newblock {\em J. Chem. Phys.}, 89(5):2965--2974, 1988.

\bibitem{plasma_sources}
H.~Conrads and M.~Schmidt.
\newblock {\em Plasma Sources Sci. T.}, 9(4):441, 2000.

\bibitem{transitions_singlet}
R.~Celiberto, M.~Capitelli, N.~Durante, and U.~T. Lamanna.
\newblock {\em Phys. Rev. A}, 54:432--438, Jul 1996.

\bibitem{experimental_vibrational_temperature}
T.~Mosbach.
\newblock {\em Plasma Sources Sci. Technol.}, 14(3):610, 2005.

\bibitem{the_simulation_paper}
M.~Capitelli et~al.
\newblock {\em Nucl. Fusion}, 46(6):S260, 2006.

\bibitem{FantzMolC}
U.~Fantz, B.~Schalk, and K.~Behringer.
\newblock {\em New J. Phys.}, 2(1):7, 2000.

\bibitem{the_book}
M.A. Lieberman and A.J. Lichtenberg.
\newblock {\em Principles of Plasma Discharges and Materials Processing}.
\newblock Wiley, 2005.

\bibitem{flat_EEDF_1}
T.~Shibata~et al.
\newblock {\em AIP Conf. Proc.}, 1515(1):177--186, 2013.

\bibitem{flat_EEDF_3}
D.~Pagano, C.~Gorse, and M.~Capitelli.
\newblock {\em IEEE Trans. Plasma Sci}, 35(5):1247--1259, 2007.

\bibitem{flat_EEDF_2}
J.~Bretagne, G.~Delouya, C.~Gorse, M.~Capitelli, and M.~Bacal.
\newblock {\em J. Phys. D: Appl. Phys.}, 18(5):811, 1985.

\bibitem{Graham_1995_kinetics_of_negative_hydrogen_ions}
W.~G. Graham.
\newblock {\em Plasma Sources Sci. Technol.}, 4(2):281, 1995.

\bibitem{Sun_2010_RF_H2_PIC_simulation}
J.~Sun, X.~Li, C.~Sang, W.~Jiang, P.~Zhang, and D.~Wang.
\newblock {\em Phys. Plasmas}, 17(10), 2010.

\bibitem{McNeely_2008_Langmuir_probe_from_batman}
P.~McNeely, S.~V Dudin, S.~Christ-Koch, U.~Fantz, and the NNBI~Team.
\newblock {\em Plasma Sources Sci. Technol.}, 18(1):014011, 2009.

\bibitem{crosssections1}
H.~Tawara, Y.~Itikawa, H.~Nishimura, and M.~Yoshino.
\newblock {\em J. Phys. Chem. Ref. Data}, 19(3):617--636, 1990.

\bibitem{crosssections2}
J.-S. Yoon, M.-Y. Song, J.-M. Han, S.~H. Hwang, W.-S. Chang, B.~Lee, and
  Y.~Itikawa.
\newblock {\em J. Phys. Chem. Ref. Data}, 37(2):913--931, 2008.

\bibitem{Celiberto_2001_cross_section_functions}
R.~Celiberto, R.~K. Janevand~A. Laricchiuta, M.~Capitelli, J.~M. Wadehra, and
  D.~E. Atems.
\newblock {\em At. Data Nucl. Data Tables}, 77(2), 2001.

\bibitem{Shakhatov_2009_Cross_section_database}
V.A. Shakhatov and Yu.A. Lebedev.
\newblock {\em High Temperature}, 49(2):257--302, 2011.

\bibitem{BC_contribution}
M.~Nishiura.
\newblock {\em Journal of Plasma and Fusion Research}, 80(9):757--762, 2004.

\bibitem{HydrogenFranckCondon}
U.~Fantz and D.~W{\"u}nderlich.
\newblock {\em At. Data Nucl. Data Tables}, 92(6):853 -- 973, 2006.

\bibitem{BC_dissociation}
H.~Abgrall, E.~Roueff, X.~Liu, and D.~E. Shemansky.
\newblock {\em Astrophysical J.}, 481(1):557, 1997.

\bibitem{electron_impact_triplet_vibrational_dependency}
R.~Celiberto, M.~Capitelli, and A.~Laricchiuta.
\newblock {\em Phys. Scripta}, 2002(T96):32, 2002.

\bibitem{Bonnie_1988_mestable_H2_in_multicusp_ion_source}
J.~H.~M. Bonnie, P.~J. Eenshuistra, and H.~J. Hopman.
\newblock {\em Phys. Rev. A}, 37:4407--4414, 1988.

\bibitem{H2_c_lifetime}
A.~B. Wedding and A.~V. Phelps.
\newblock {\em J. Chem. Phys.}, 89(5):2965--2974, 1988.

\bibitem{LymanWernerExperiment}
J.~M. Ajello, S.~K. Srivastava, and Yuk~L. Yung.
\newblock {\em Phys. Rev. A}, 25:2485--2498, May 1982.

\bibitem{Liu_2003_Electron_impact_excitation_of_H2_EF}
X.~Liu, D~E Shemansky, H~Abgrall, E~Roueff, S~M Ahmed, and J~M Ajello.
\newblock {\em J Phys. B - At. Mol. Opt}, 36(2):173, 2003.

\bibitem{TPR010}
{O}perating manual of {B}alzers {TPR} 010 pirani gauge.

\bibitem{TRIUMFCharac}
Y.~S. Hwang~et al.
\newblock {\em Rev. Sci. Instrum.}, 77(3):03A509, 2006.

\bibitem{multicusp_confinement_1}
K.N. Leung, T.K. Samec, and A.~Lamm.
\newblock {\em Physics Letters A}, 51(8):490 -- 492, 1975.

\bibitem{nonMaxwelliansimulation1}
G.~Lombardi, X.~Duten, K.~Hassouni, M.~Capitelli, and A.~Gicquel.
\newblock {\em Eur Phys J. D.}, 30(2):225--233, 2004.

\bibitem{metastable_helium}
K.~Baldwin.
\newblock {\em Contemp. Phys.}, 46(2):105--120, 2005.

\end{thebibliography}

\end{document}